\documentclass[review]{elsarticle}

\usepackage{hyperref}
\usepackage{graphicx}

\usepackage{xcolor}
\usepackage{amsmath,amssymb}

\usepackage{tikz}
\usetikzlibrary{decorations.pathmorphing,shapes}

\journal{Physica D: Nonlinear Phenomena}

\begin{document}

\begin{frontmatter}

\title{A shadowing-based inflation scheme for ensemble data assimilation}

\author{Thomas Bellsky}%\corref{mycorrespondingauthor}
\address{Department of Mathematics and Statistics, University of Maine, Orono, ME, USA}

\author{Lewis Mitchell\corref{mycorrespondingauthor}}%\corref{mycorrespondingauthor}
\address{School of Mathematical Sciences, University of Adelaide, Adelaide, South Australia, Australia}
\cortext[mycorrespondingauthor]{Corresponding author. {\it Telephone}: +61 8 8313 5424 {\it Facsimile}: +61 8 8313 3696}
\ead{lewis.mitchell@adelaide.edu.au }
\ead[url]{http://maths.adelaide.edu.au/lewis.mitchell}

\begin{abstract}
Artificial ensemble inflation is a common technique in ensemble data assimilation,
whereby the ensemble covariance is periodically increased in order to prevent deviation of the ensemble from the observations and possible ensemble collapse. 
This manuscript introduces a new form of covariance inflation for ensemble data assimilation based upon shadowing ideas from dynamical systems theory.
%We present results from a low order nonlinear chaotic system that supports using shadowing inflation, demonstrating that shadowing inflation outperforms standard multiplicative covariance inflation in observation-sparse scenarios and is more robust to parameter tuning.
We present results from a low order nonlinear chaotic system that support using shadowing inflation, demonstrating that shadowing inflation is more robust to parameter tuning than standard multiplicative covariance inflation, often leading to longer forecast shadowing times.
%outperforming in observation-sparse scenarios and 
\end{abstract}


\end{frontmatter}

%\linenumbers
\newpage

\section{Introduction}

Ensemble filtering methods are Monte Carlo approximations to the Bayesian update problem of combining a prior probability density function for a system state with a likelihood function for observational data.
Such methods are widely used across the geophysical sciences,
for improving forecasts in numerical weather and climate prediction \cite{houtekamer2016review, Rabier2005}
as well as in ocean \cite{evensen1994inverse, Vernieres2011} and atmospheric science \cite{anderson2007exploring, Kalnay2002}.
A key advantage of using ensemble methods is the approximation of distributions by finite-size ensembles,
leading to a massive computational advantage and the ability to represent otherwise inaccessible high-dimensional distributions \cite{houtekamer2001sequential}.
The ensemble Kalman filter (EnKF) \cite{Evensen} along with its variants and extensions such as the local ensemble transform Kalman filter (LETKF) \cite{LETKF} are efficient data assimilation methods which have proven highly effective across a range of applications involving both state and parameter estimation
\cite{delsole2010state, Bellsky2014a, Bellsky2014, kretschmer2015}.

Ensemble approximations come at an important cost:
the use of finite-size ensembles generically leads to sampling errors in ensemble-based Kalman filtering techniques,
which can manifest in a number of ways.
The most common issue related to insufficient ensemble size is misspecification of error covariances.
Analysis ensemble spreads are often routinely underestimated \cite{li2009simultaneous},
or possibly overestimated, 
particularly in sparse observational grids \cite{Liu2008,Whitaker2009,Gottwald2011}.
Covariance underestimation can lead to filter divergence where the analysis ensemble becomes overconfident in model forecasts and fails to track the true system states or observations,
in some cases even leading to numerical instabilities in the forecast model which ultimately catastrophically diverge to machine infinity \cite{kelly2015concrete, Harlim2010}.
A mechanism for such filter divergence is a finite-size ensemble aligning away from a sufficiently strong attractor, 
which can lead to integrating a stiff dynamical system \cite{Gottwald2013}.

One remedy to counter the underestimation of the error covariance and possible filter divergence is to artificially inflate the ensemble covariance. 
This artificial inflation can be done simply by periodically adding noise or applying a multiplicative factor greater than one to the error covariance \cite{evensen2009data, burgers1998analysis}. There also exist adaptive and hybrid covariance inflation methods which modify the error covariance by accounting for various features of the forecast model and ensemble \cite{mitchell2000adaptive, anderson2007adaptive,anderson2009spatially,miyoshi2011gaussian}. Additionally, there exist hybrid methods avoiding inflation altogether when the analysis strongly dominates the prior \cite{bocquet2015expanding}. Other methods to prevent filter divergence involve modifying the basic ensemble filtering algorithms through judicious stochastic parameterization \cite{Harlim2010a,Mitchell2012}. 
Finite ensemble sizes can also lead to spurious correlations appearing in the error covariance matrices. 
Such spurious correlations may be ameliorated by spatially localizing the effects of observations \cite{anderson2007exploring}.

Ideally, a forecast will remain close to the true state or observations for as long as possible
and not exhibit any form of divergence,
catastrophic or otherwise.
Mathematical shadowing theory,
developed in the context of hyperbolic systems \cite{Anosov1967,Bowen1975},
provides rigorous results guaranteeing the existence of true model trajectories which remain close to a given pseudo-trajectory 
(one which is almost an actual trajectory of the model) 
for arbitrarily long times. 
Such theory has been advanced to show that numerical solutions of chaotic systems do indeed approximate true trajectories \cite{hammel1987numerical, hammel1988numerical}. 
Other studies have determined methods of numerical approximation
for hyperbolic periodic orbits that shadow actual periodic orbits \cite{van1995numerical}. 
Of course, for non-hyperbolic systems it can be proven that no such shadowing trajectory exists (e.g. \cite{Kostelich1997}),
in which case the aim becomes to find shadowing trajectories which nonetheless remain close to pseudo-trajectories with only small mismatches \cite{Judd2008}.

Operationally, finding shadowing trajectories is greatly limited by 
model error,
confounding sources of error from observations,
as well as the sparsity of observations.
Nonetheless,
\cite{Danforth2006} proposed a simple method, 
later extended in \cite{Lieb-Lappen2012},
for inflating ensemble forecasts in a non-hyperbolic chaotic dynamical system. Their method, which only inflates the ensemble in directions in which uncertainty is shrinking, has had success in increasing the shadowing time of a model forecast.
While this method of course does not guarantee the existence of shadowing trajectories in the dynamical systems sense,
it has nonetheless shown favourable numerical results in low-dimensional chaotic dynamical systems,
suggesting a potential application to data assimilation problems.

In this article we introduce a new covariance inflation method for ensemble data assimilation 
which addresses the general problem of filter divergence using the shadowing-based approach from \cite{Danforth2006, Lieb-Lappen2012}. Our method aims to judiciously inflate the ensemble only in directions in which the ensemble is growing overconfident, 
with the intention of keeping the analysis ensemble close to the attractor.
This is done through an algorithm that identifies the contracting ensemble directions over the forecast cycle,
and then inflates the ensemble only in these contracting directions before performing the DA analysis.
We apply our shadowing inflation method within the context of the LETKF,
and compare it with the standard multiplicative inflation method via numerical twin experiments 
on a low-dimensional non-hyperbolic nonlinear chaotic system exhibiting dynamics akin to those in the atmosphere.
As we will demonstrate through numerical simulations, 
the proposed approach works well in comparison to standard multiplicative methods, in that shadowing inflation is less sensitive to inflation parameter tuning, reduces error and dispersion, while maintaining ensemble reliability.

The remainder of this paper is organized as follows: Section 2 describes ensemble data assimilation and covariance inflation, 
and proposes the shadowing inflation algorithm used to ameliorate issues related to covariance underestimation.
Section 3 details the model and setup of our numerical experiments,
and Section 4 presents numerical results comparing standard multiplicative inflation to our proposed shadowing inflation method.
We conclude with a discussion in Section 5.

\section{Ensemble data assimilation}

A general ensemble Kalman filter is a Monte Carlo data assimilation technique based on the Kalman filter \cite{Kalman}. The Kalman filter is an algorithm for determining a state estimate using both a model prediction and observational data. An ensemble Kalman filter is particularly useful, because it extends the linear Kalman filter to nonlinear models and is computational efficient for large state space vectors \cite{LETKF}. 

To mathematically describe this type of filter, we assume we have some forecast model $M$ that sequentially determines a model state $z$. The model advances the previous analysis state to the forecast state $z^f$, which at time $t_j$ is:
\begin{equation}
z_{j}^{f(i)} =M\left(z_{j-1}^{a(i)}\right).
\end{equation}
Thus, $M$ takes the previous analysis state estimate and updates it forward in time. The index $i$ notates a particular ensemble state, where there are $k$ ensemble states:
\begin{equation}
\left\{z_{j-1}^{a(i)} : i=1,2,\dots,k\right\}.
\end{equation}
Furthermore, there exist spatial observations of the state at time $t_j$, denoted by the vector $y$. Typically, the number of observations is much less than the size of the state space $\dim (y) \ll \dim(z)$. In this formulation, it is assumed that there is a linear observation operator $H$ that projects the state space to the observation space:
\begin{equation}
y_j=H z_j+\epsilon_j.
\end{equation}
The $z_j$ above represents the true state at time $t_j$ and the observational error is assumed to be a Gaussian random variable $\epsilon_j  \sim N \left(0,R_j\right)$, where $R_j$ is the covariance matrix for the observations. For simplicity, the time-step notation $j$ will be dropped in the forthcoming notation.

The ensemble Kalman filter \cite{Evensen, houtekamer1998data} is a reduced rank filter, where an ensemble of $k$ analysis states from the previous time step $Z^a = \left\{z_1^a,z_2^a, \dots z_k^a\right\}$ is each individually advanced forward by the forecast model to determine the background forecast ensemble $Z^f = \left\{z_1^f,z_2^f, \dots z_k^f\right\}$. Then the background forecast covariance is formulated as:
\begin{equation}
\label{CovPb} P^f = \frac{1}{k-1}Z^{f'}(Z^{f'})^T,
\end{equation}
where the $i$-th column of $Z^{f'}$ is $z_{i}^{f}-\bar z^f$, with $\bar z^f$ indicating the mean $\bar z^f = \frac{1}{k}\sum_{i=1}^k z_i^{f}$. Thus, the background forecast covariance \eqref{CovPb} is not invertible since it is of rank $k-1$, so various ensemble Kalman filter methods perform a change of coordinates to determine a Kalman filter update step, where transform methods avoid computing the  background forecast covariance altogether. In particular, the local ensemble transform Kalman filter (LETKF) \cite{LETKF} updates the covariance analysis as:
\begin{equation}
\label{CovPa} P^a = Z^{f'}\tilde P^a \left(Z^{f'}\right)^T,
\end{equation}
where $\tilde P^a$ represents the analysis error covariance in a $k$ dimensional space:
\begin{equation}
\label{tildaPa} \tilde P^a = \left[(k-1)I+(HZ^{f'})^T R^{-1}\left(Z^{f'}\right)\right],
\end{equation}
and updates each ensemble state as:
\begin{equation}
\label{za} z_i^a = \bar z^f + Z^{f'} \tilde P^a \left(H Z^{f'} \right)^TR^{-1}\left(y-Hz_i^f\right).
\end{equation}

In this study, we use the LETKF as our data assimilation method. The LETKF makes use of localization, where a state location is updated by only considering nearby observations. A basic technique for performing localization is to assign some universal localization radius $r$, and then only update a state location using observations within $r$ units of that location.

\subsection{Covariance inflation}

A particular issue with ensemble Kalman filters is that ensemble states often tend to the ensemble mean with small uncertainty. This can lead to the problem of ensemble collapse, where the EnKF analysis leads to an overconfident, but incorrect state, no longer shadowing the truth.

Ensemble covariance inflation is a procedure to avoid underestimating uncertainties and ensemble collapse. These methods artificially inflate uncertainties in the background covariance. As discussed in the introduction, there are a variety of techniques for performing covariance inflation. One common method is multiplicative inflation, where the background forecast covariance is inflated by a multiplicative factor $1+\delta$ for $\delta >0$, thus $P^f \rightarrow \left(1+\delta\right) P^f.$ A similar technique is additive inflation, which adds noise $\Upsilon$ to the background covariance $P^f  \rightarrow P^f + \Upsilon.$ 

\subsection{Shadowing inflation}

We present a new type of covariance inflation, based on ideas from \cite{Lieb-Lappen2012}.
That work examines how long a forecast ensemble ``shadows'' a true solution, but the techniques do not involve any observations or data assimilation. 
As discussed in the introduction, when an ensemble shadows the true solution, this can be described as the spread of the ensemble (cloud of uncertainty) containing the true solution. 
Typically, one might begin with a well-distributed ensemble of state solutions $Z_0 \in \mathbb R^{N \times k}$ made up of $k$ ensemble members. 
Each ensemble member is propagated forward by the forecast model $M$, and a singular value decomposition (SVD) is performed at each evaluation time step $t_j$:
\begin{equation}
Z_j= U_j S_j V_j^T,
\end{equation}
where the resulting singular values (the diagonal components of $S_j$)  determine the length of the axes of ensemble spread and the singular vector $U_j$ determines the direction. 
The SVD approximates the region of uncertainty after model propagation,
with expanding directions of uncertainty stretching the ellipse and collapsing directions of uncertainty shrinking the ellipse, 
which is well described in \cite{Danforth2006}.
We remark that in reality the SVD provides a linear approximation to the true ellipse of uncertainty created by the nonlinear evolution of the forecast dynamics, 
however for short forecast intervals this difference often remains small.

\begin{figure}
\begin{center}

\begin{tabular}{lr}

\begin{tikzpicture}

\draw (-0.8,3.5) node [anchor=west] {(a) Standard};

% true trajectory
\draw [thick] (0,0) .. controls (1,3) and (2,1.5) .. (3.5,3);
\draw (1.9,2.1) node [anchor=south] {$z^t$};

% initial epsilon-ball
\filldraw (0,0) circle (3pt);
\filldraw [opacity=0.4] (0,0) circle (15pt);
%\draw [->,thick] (0,0) -- (-0.37,0.37);
%\draw (-0.09,0.09) node [anchor=northeast] {$\varepsilon$};
\draw (0,-.07) node [anchor=north] {$Z_0$};

% final epsilon-ball
\filldraw (3.5,3) circle (3pt);
\filldraw [opacity=0.4] (3.5,3) circle (15pt);

% forecast
\draw [dashed] (0,0.53) .. controls (1.75,0.5) and (2,1.9) .. (3.5,2.5);
\draw [dashed] (0,-0.53) .. controls (3,-0.2) and (2.5,1.25) .. (3.8,1.85);
\filldraw (3.5,2.1) [opacity=0.4,red,rotate around={30:(3.5,2.1)}] ellipse (30pt and 10pt);
% \draw (3.8,1) .. controls (4,1.2) and (3.5,2) .. ;

% inflation
\filldraw (3.5,2.1) [opacity=0.4,red,rotate around={30:(3.5,2.1)}] ellipse (42pt and 14pt);
\draw (4.2,1.3) node [anchor=north] {$\left(1+\delta\right) Z^f$};
\draw (4,1.3) .. controls (3.9,2.2) and (4.6,1.8) .. (4.6,2.7);
\filldraw (4.6,2.7) circle (1pt);
\draw (3.2,1.9) node {$Z^f$};

% analysis
\filldraw (3.5,3.5) [opacity=0.6,red,rotate around={20:(3.5,3.5)}] ellipse (21pt and 8pt);
%\draw (2.2,3.4) .. controls (2.4,3.2) and (2.7,3.6) .. (3.1,3.4);
%\draw (2.2,3.4) 
\draw (2.1,3.2) .. controls (2.4,3.1) and (2.7,3.4) .. (3,3.4);
\draw (2.1,3.2) node [anchor=east] {$Z^a$};
\filldraw (3,3.4) circle (1pt);

% obs
\draw (3.5,3.6) node {\Large $\ast$};
\draw (3.5,3.65) node [anchor=south] {$y$};

\end{tikzpicture}

&

\begin{tikzpicture}

\draw (-0.8,3.5) node [anchor=west] {(b) Shadowing};

% true trajectory
\draw [thick] (0,0) .. controls (1,3) and (2,1.5) .. (3.5,3);
\draw (1.9,2.1) node [anchor=south] {$z^t$};

% initial epsilon-ball
\filldraw (0,0) circle (3pt);
\filldraw [opacity=0.4] (0,0) circle (15pt);
%\draw [->,thick] (0,0) -- (-0.37,0.37);
%\draw (-0.09,0.09) node [anchor=northeast] {$\varepsilon$};
\draw (0,-.07) node [anchor=north] {$Z_0$};

% final epsilon-ball
\filldraw (3.5,3) circle (3pt);
\filldraw [opacity=0.4] (3.5,3) circle (15pt);

% forecast
\draw [dashed] (0,0.53) .. controls (1.75,0.5) and (2,1.9) .. (3.5,2.5);
\draw [dashed] (0,-0.53) .. controls (3,-0.2) and (2.5,1.25) .. (3.8,1.85);
\filldraw (3.5,2.1) [opacity=0.4,blue,rotate around={30:(3.5,2.1)}] ellipse (30pt and 10pt);
% \draw (3.8,1) .. controls (4,1.2) and (3.5,2) .. ;

% inflation
\filldraw (3.5,2.1) [opacity=0.4,blue,rotate around={30:(3.5,2.1)}] ellipse (30pt and 17pt);
\draw (3.5,0.3) node [anchor=north] {$\bar z^f I+\mathcal{M}Z^{f'}$};
\draw (3.5,0.3) .. controls (3.3,1.0) and (4.3,1.6) .. (3.9,1.75);
\filldraw (3.9,1.75) circle (1pt);
\draw (3.2,1.9) node {$Z^f$};

% analysis
\filldraw (3.5,3.4) [opacity=0.5,blue,rotate around={20:(3.5,3.5)}] ellipse (15pt and 10pt);
\draw (2.5,3.3) .. controls (2.5,3.3) and (2.9,3.2) .. (3.15,3.45);
\draw (2.5,3.3) node [anchor=east] {$Z^a$};
\filldraw (3.15,3.45) circle (1pt);

% obs
\draw (3.5,3.6) node {\Large $ \ast$};
\draw (3.5,3.65) node [anchor=south] {$y$};

\end{tikzpicture}

\end{tabular}

\caption{The two figures above are two dimensional cartoons illustrating standard multiplicative inflation and shadowing inflation. In both, there exists an uncertainty for the true trajectory $z^t$, where the initial data's uncertainty  $Z_0$ is well-distributed about the initial condition. 
In (a) the forecast model carries forward the ensemble of trajectories to a future time, 
where some overlap occurs between the actual uncertainty $Z^t$ and the final analysis uncertainty $Z^a$,
after the forecast ensemble $Z^f$ has been inflated by a factor $1+\delta$ and the observation $y$ has been assimilated.
In (b) the proposed shadowing inflation scheme is illustrated, 
where only the shrinking dimension in the uncertainty $Z^f$ is inflated after the forecast, leading to the analysis ensemble $Z^a$.
The shadowing inflation scheme often leads to a greater overlap between the analysis ensemble and the true uncertainty, with subsequent forecasts achieving a longer shadowing time.
\label{fig:schematic}
}
\end{center}
\end{figure}
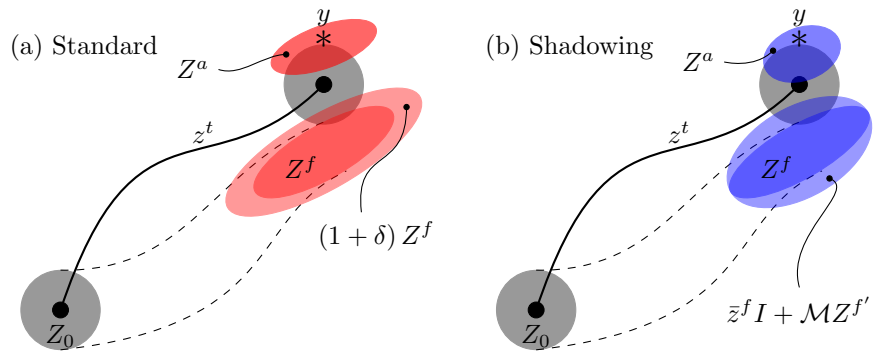

In \cite{Lieb-Lappen2012}, the concept of stalking, an aggressive form of shadowing, is introduced. Under stalking, at each evaluation time step, artificial uncertainty is inserted in the shrinking directions of the ellipse. Their results determined that the stalking methodology often led to ensembles shadowing the truth for a longer forecast period.

We adapt the idea of ensemble stalking for the purpose of ensemble inflation, 
which we call \emph{shadowing inflation}.
We remark that while \emph{shadowing} also has a technical definition within dynamical systems theory,
we use the term here specifically in reference to the method developed in \cite{Lieb-Lappen2012}.
At an assimilation time step $t$, the forecast model has determined the forecasted ensemble background state $Z^f(t)=Z^f \in \mathbb R^{N \times k}$. From this forecasted ensemble state, the shadowing inflation scheme performs the following steps:

1. Form the matrix: $Z^{f'}(t)=Z^{f'} \in \mathbb R^{N \times k}$ (recall the $i$-th column of $Z^{f'} $ is $z^{f(i)} - \bar z^f$) at the beginning of some DA analysis time step $t$. From a recent, but previous, model step $t_-$ (occurring after the last DA cycle), we also have a forecasted ensemble background state $Z^{f}(t_-)=Z^{f}_{-}$, from which we similarly form: $Z^{f'}(t_{-})=Z^{f'_-} \in \mathbb R^{N \times k}.$ Here, we are implicitly assuming there are multiple forecasts steps between each DA cycle. For instance, a numerical weather forecast is typically advanced over many incremental time intervals during the 6 hour period between a DA cycle.

2. Perform a singular value decomposition on both: $Z^{f'}=USV^T$ and  $Z^{f'_-}=U_{-}S_{-}V_{-}^T$ (here $S$ consists of up to $k-1$ nonzero singular values $s_i(t)$ and $S_{-}$ consists of up to $k-1$ nonzero singular values $s_i(t_-)$). 
The length and direction of the ensemble ellipsoid (spread) will be determined by $s_iu_i$, where the $u_i$'s are the columns of $U$.
Due to the non-uniqueness of the SVD, the directions of the singular vectors in $U_{-}$ need to be matched with their corresponding vectors in $U$.
We do this by calculating the absolute value of the dot products between all pairs of vectors,
and then selecting the set of pairs with maximal absolute values.

3. Determine the columns $u_c$ of a new matrix $U_c$ by determining all $i$ for which: $u_c=\left\{ u_i: s_i(t) < s_i(t_-)\right\}$

4. Form the inflation matrix: 
\begin{equation}
\mathcal{M} = I+ \delta U_cU_c^T; \label{inflationmatrix}
\end{equation}
where $\delta>0$ is a (small) constant inflation parameter.

5. Finally, form the shadowing inflated background ensemble state: $Z^b(t) =\bar z^f I+\mathcal{M}Z^{f'},$ from which the data assimilation process is continued to determine the analysis state.

A schematic diagram illustrating the method plus the analysis step is given in Figure \ref{fig:schematic}.
This process only inflates the contracting eigendirections, and shadowing inflation only acts in the directions spanned by the analysis ensemble.
It performs no inflation on the expanding eigendirections, however this could be easily incorporated (as could a deflation in these directions) if desired. 
In the numerical experiments which follow we found that inflating the expanding eigendirections was detrimental to the analysis.

\section{Model and experimental setup}

We use the Lorenz-96 model \cite{Lorenz1996} as a test bed for our experiments and results. Lorenz-96 is a conceptual model that determines a `weather' state on a latitude circle: 
\begin{equation*}
\frac{d z_i}{d t} = \left(z_{i+1} +z_{i-2}\right)z_{i-1} - z_i + F.
%\quad X_{i\pm N} =& X_i.
\end{equation*}
In this model, the nonlinear terms mimic advection and conserve the total energy. The linear term dissipates the total energy. $F$ is the forcing, which strongly determines chaotic properties. For our experiments, we take $N=40$ locations on this latitude circle. We assume the standard forcing $F=8$, which corresponds to a chaoticity similar to true atmospheric dynamics \cite{Lorenz1996}.
The climatological standard deviation for the system with these parameters is $\sigma_{\rm clim} = 3.63$.

For this model and choice of parameters, a time-step of $h=0.05$ simulates a 6 hour Earth weather forecast \cite{Lorenz1996}. We discretize this model on a $h/10$ time-step, performing DA updates at multiples of $h$. When performing shadowing inflation, we determine the expanding (and contracting) directions of uncertainty by examining the singular value decomposition at the assimilation time step $t_a$ and the previous model step $t_a-h/10$.

We run the model for $110$ days of model time ($t = 22$),
create $N_{\rm obs}$ equally-spaced, fixed, synthetic observations by adding Gaussian noise with error covariance $R = 0.2 I$ to every observed spatial location,
and allow the system to spin up for 10 days from a random initial condition before performing experiments.
We perform 100 separate simulations, where the majority of our results provide the median of the RMS error and the corresponding interquartile range of RMS error over the 100 simulations. We focus on the median and interquartile range of our sample of simulations here due to the skewed nature of the sample -- a small number of initial condition and observation sets lead to filter divergence with both forms of inflation,
which dominate the calculation of mean RMS errors and obscure the trends observed. 
We use $k = 20$ ensemble members in all experiments, randomly initialized by adding Gaussian noise with error covariance equal in size to that of the observations to the truth at $t = 0$.
In all experiments we set the localization radius $r = 5$,  which we found to produce the best performance in the LETKF with global multiplicative inflation, and set  $N_{\rm obs} = 8$. We vary our constant inflation parameter, in Equation \ref{inflationmatrix}, anywhere from $\delta=0.005$ to $\delta=0.1$ (which is equivalent to a matched fixed multiplicative inflation factor varying from of $1.005$ to $1.1$)

\section{Results}
\begin{figure}
\begin{center}
\includegraphics[width=\columnwidth]{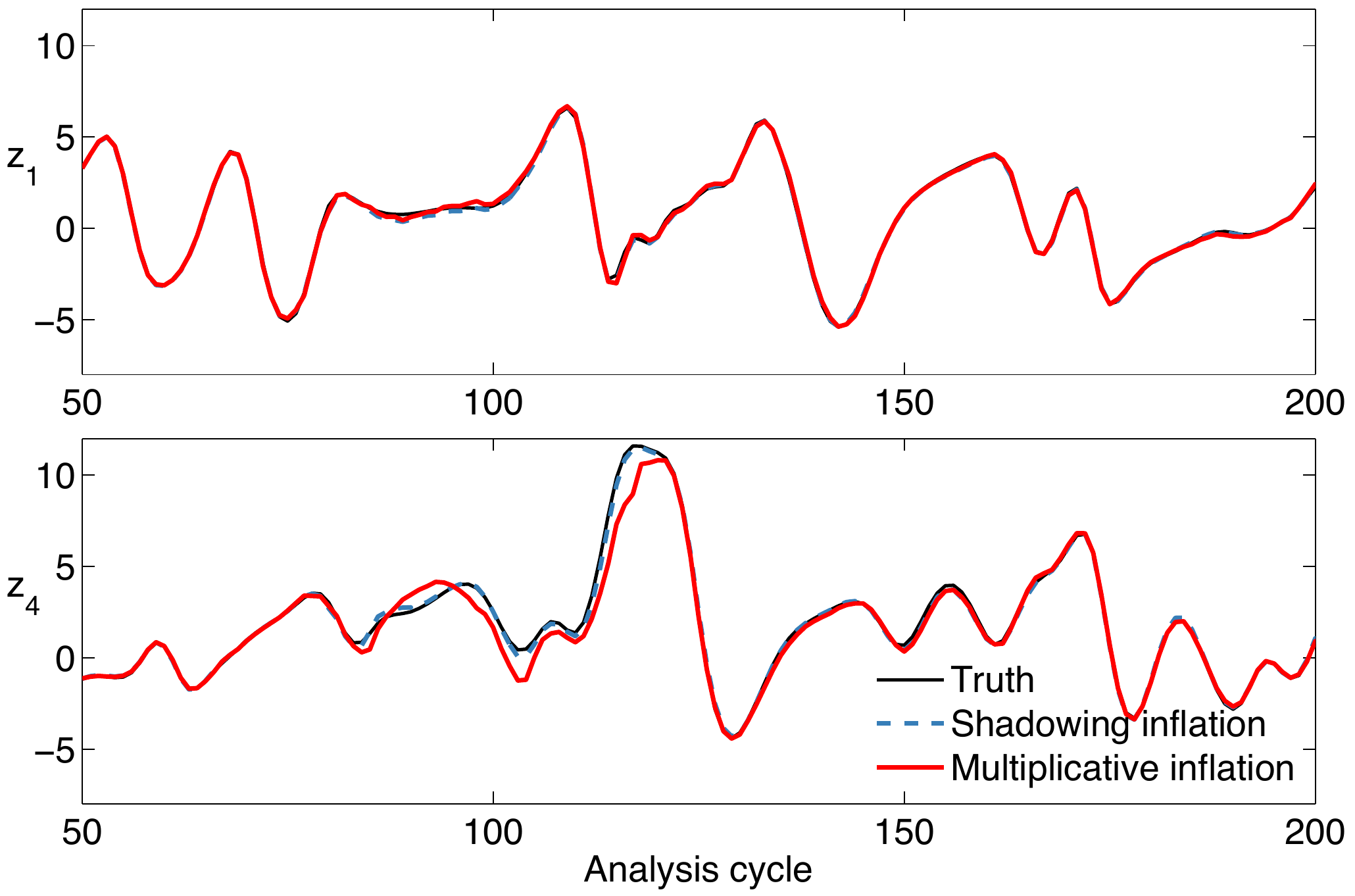}
\end{center}
\caption{Example analysis trajectories for standard (red, solid) and shadowing (blue, dashed) methods. Top: observed component. Bottom: unobserved component. Black, solid line shows the truth.
}
\label{fig:sampleTraj}
\end{figure}

In Figure \ref{fig:sampleTraj} we show example analyses made by
the standard multiplicative inflation and shadowing inflation methods with a shadowing inflation parameter of $\delta=0.02$ (and multiplicative inflation factor of $1.02$),
for an observed and unobserved component of Lorenz-96.
While both inflation methods track the observed component equally well, the
shadowing inflation tracks the unobserved component of the truth more closely,
especially between analysis cycles 85 and 120,
where the standard method diverges noticeably from the truth.

\begin{figure}
\begin{center}
\includegraphics[width=\columnwidth]{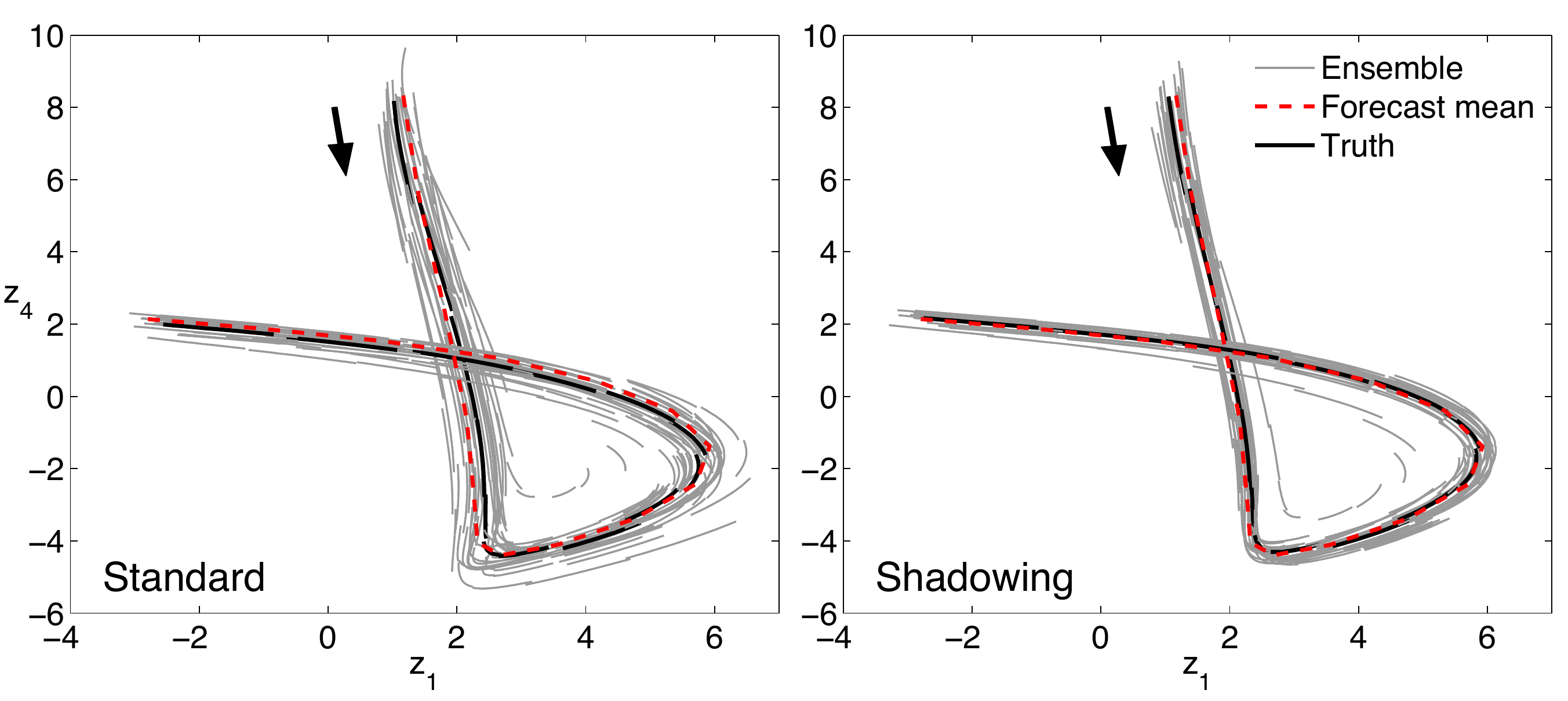}
\end{center}
\caption{Example ensemble trajectories for standard (left) and shadowing (right) inflation schemes.
The grey trajectories show forecasts generated by individual ensemble members after each analysis,
while the red dashed line shows the mean of these ensemble forecasts.
The solid black line shows the true trajectory being estimated.}
\label{fig:sampleEnsembleTraj}
\end{figure}

We focus more on this improvement shadowing inflation makes over the standard scheme in Figure \ref{fig:sampleEnsembleTraj},
where for the same parameter values as in Figure \ref{fig:sampleTraj}, we show individual ensemble member trajectories between analysis cycles 120 and 140.
Throughout, the standard inflation scheme greatly overinflates the ensemble,
leading to an analysis ensemble which is broadly dispersed around the true trajectory.
This is particularly noticeable where the true trajectory passes through local extrema (turning points for the $z_1$ and $z_4$ coordinates shown),
where the spread around the truth appears relatively large.
On the other hand, the shadowing inflation scheme avoids this overinflation by only inflating the ensemble in contracting directions,
leading to an ensemble which is more tightly clustered around the truth throughout.
This suggests that shadowing inflation is most beneficial near the edges of the model attractor,
similar to results in \cite{Lieb-Lappen2012} for the forecasting problem.
We might also expect shadowing inflation to outperform the standard method near the stable manifold of a saddle, 
where trajectories in the analysis ensemble might diverge dramatically due to being falsely initialized on both sides of the manifold.
Note also that while the analysis mean for each ensemble is not visibly dissimilar in our example
(Figure \ref{fig:sampleTraj}, analysis cycles 120--140),
there is a substantial difference in how well each ensemble represents the truth.
Indeed, in Figure \ref{fig:sampleEnsembleTraj} the forecast mean (red dashed line) in this region is reasonably close to the truth and not dissimilar between the two inflation methods,
however the ensemble spread is noticeably worse for the global multiplicative inflation.

\begin{figure}
\begin{center}
\includegraphics[width=.95\columnwidth]{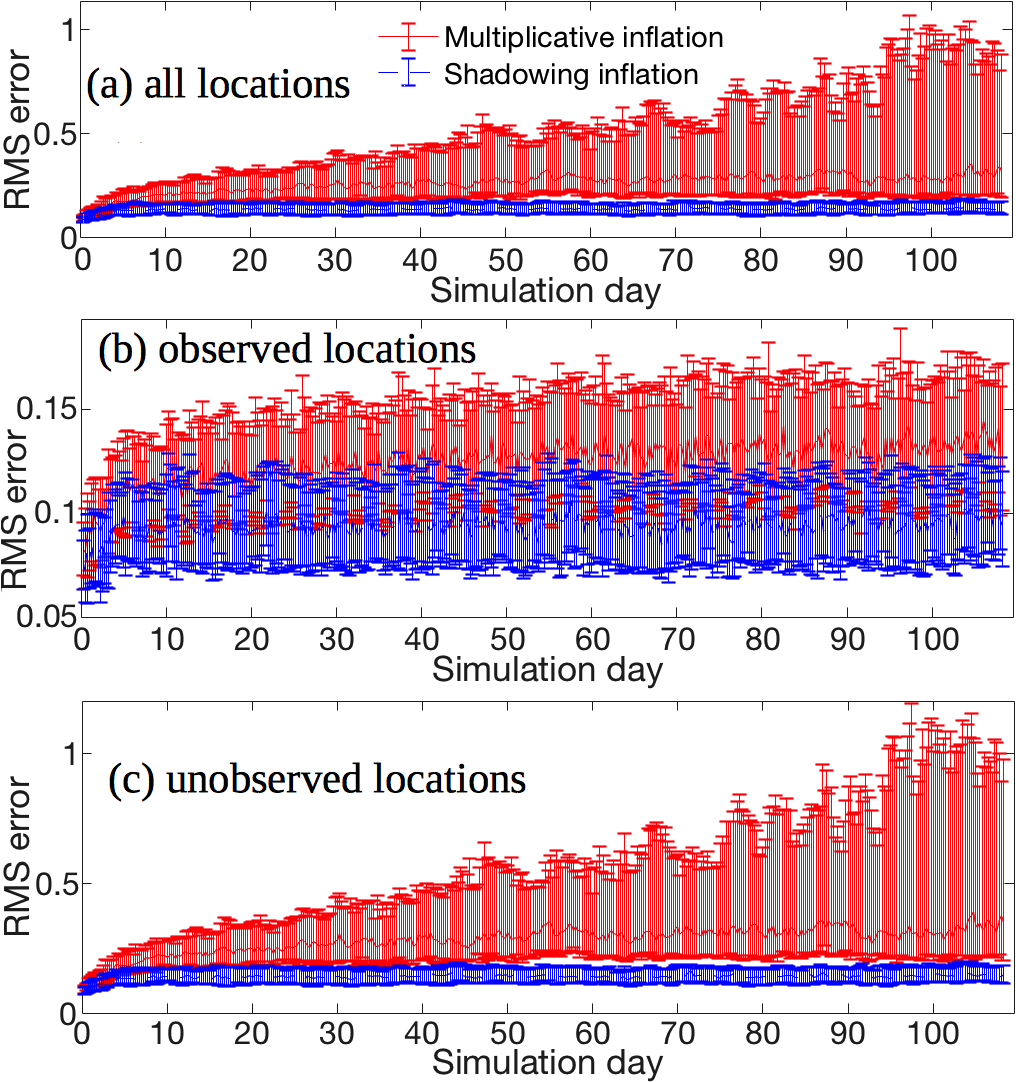}
\label{fig:timeseries}
\end{center}
\caption{The median and interquartile RMS errors for the multiplicative inflation scheme (red, solid) and shadowing inflation scheme (blue, dashed) are plotted over the entire simulation time (with a shadowing inflation parameter $\delta=0.05$ and multiplicative inflation $1.05$). Subplots show errors averaged over (a) all locations; (b) observed locations; (c) unobserved locations. For comparison, the climatological standard deviation is $\sigma_{\rm clim} = 3.63$ and observational error standard deviation is $\sqrt{0.2} \approx 0.44$ }
\end{figure}

Figure \ref{fig:sampleTraj} suggests that the shadowing inflation scheme is most beneficial in the unobserved subspace.
To explore this we plot the median and interquartile RMS errors at each DA step over the entire 110 day simulation (here $\delta=0.05$ and multiplicative inflation $1.05$), first for all state locations in Figure  \ref{fig:timeseries}a, as well as the median and interquartile RMS errors in only the observed and unobserved state subspaces (Figures \ref{fig:timeseries}b and \ref{fig:timeseries}c respectively).

\begin{figure}
\begin{center}
\includegraphics[scale=.28]{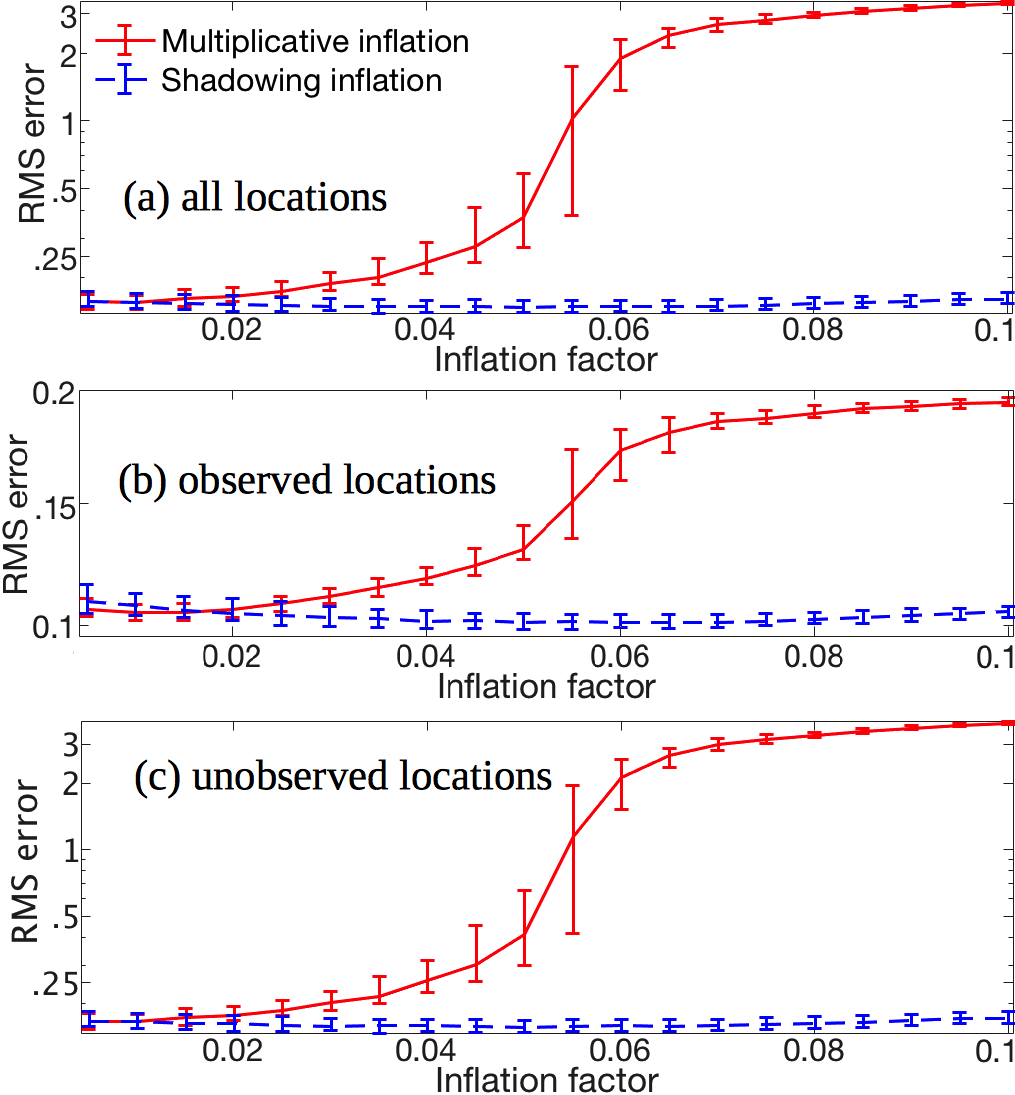}
\end{center}
\caption{Top three figures above plot the median and interquartile RMS errors of the multiplicative inflation scheme  (solid, red) and the shadowing inflation scheme (dashed, blue), varying the inflation factor. (a) is a log scale of the RMS error of all locations, (b) is the RMS error of observed locations, and (c) is a log scale of the RMS error of unobserved locations. Here, the localization radius is $r=5$ and there are $8$ fixed observations.}
\label{Compare_Infl}
\end{figure}

Examining all locations in Figure  \ref{fig:timeseries}a, the median and interquartile RMS errors of the shadowing method are relatively small, where the median RMS error of the multiplicative method is larger with the RMS error in the upper quartile increasing in forward time. The RMS errors of the observed locations in Figures \ref{fig:timeseries}b are both quite small, with the shadowing method doing incrementally better (due to sensitivity to the inflation parameter, further examined in Figure \ref{Compare_Infl}). Similar to Figure \ref{fig:sampleTraj}, in Figure \ref{fig:timeseries}c the shadowing inflation method significantly outperforms multiplicative inflation at unobserved locations, with a smaller median RMS error and less dispersion, especially in the upper quartile.

To compare the robustness of the shadowing and standard inflation schemes further, Figure \ref{Compare_Infl} explores the sensitivity of the two methods to the choice of inflation parameter $\delta$. This figure varies the shadowing inflation factor $\delta$ from $0.005$ to $0.1$, and equivalently varies the fixed multiplicative inflation factor $1+\delta$ from $1.005$ to $1.1$. Figures \ref{Compare_Infl}a-c respectively plot the median and interquartile RMS errors of the multiplicative inflation scheme (solid, red) and the shadowing inflation scheme  (dashed, blue) for a) all locations (log scale), b) only observed locations, c) only unobserved locations (log scale). The optimal multiplicative inflation parameter is near $\delta=0.01$ and the optimal shadowing inflation parameter is near $\delta=0.05$. We see that the shadowing inflation scheme is much more robust to the predetermined inflation factor than the multiplicative scheme, which increases in error for increasing values of $\delta$ past $0.01$. 

%Figure \ref{Compare_Infl}d plots the mean of the trace of the analysis covariance $P^a$ over all time steps, effectively showing that the shadowing scheme is again less sensitive to the inflation factor than the multiplicative scheme, which overestimates the analysis uncertainty for larger inflation factors.

\begin{figure}
\begin{center}
\includegraphics[scale=.30]{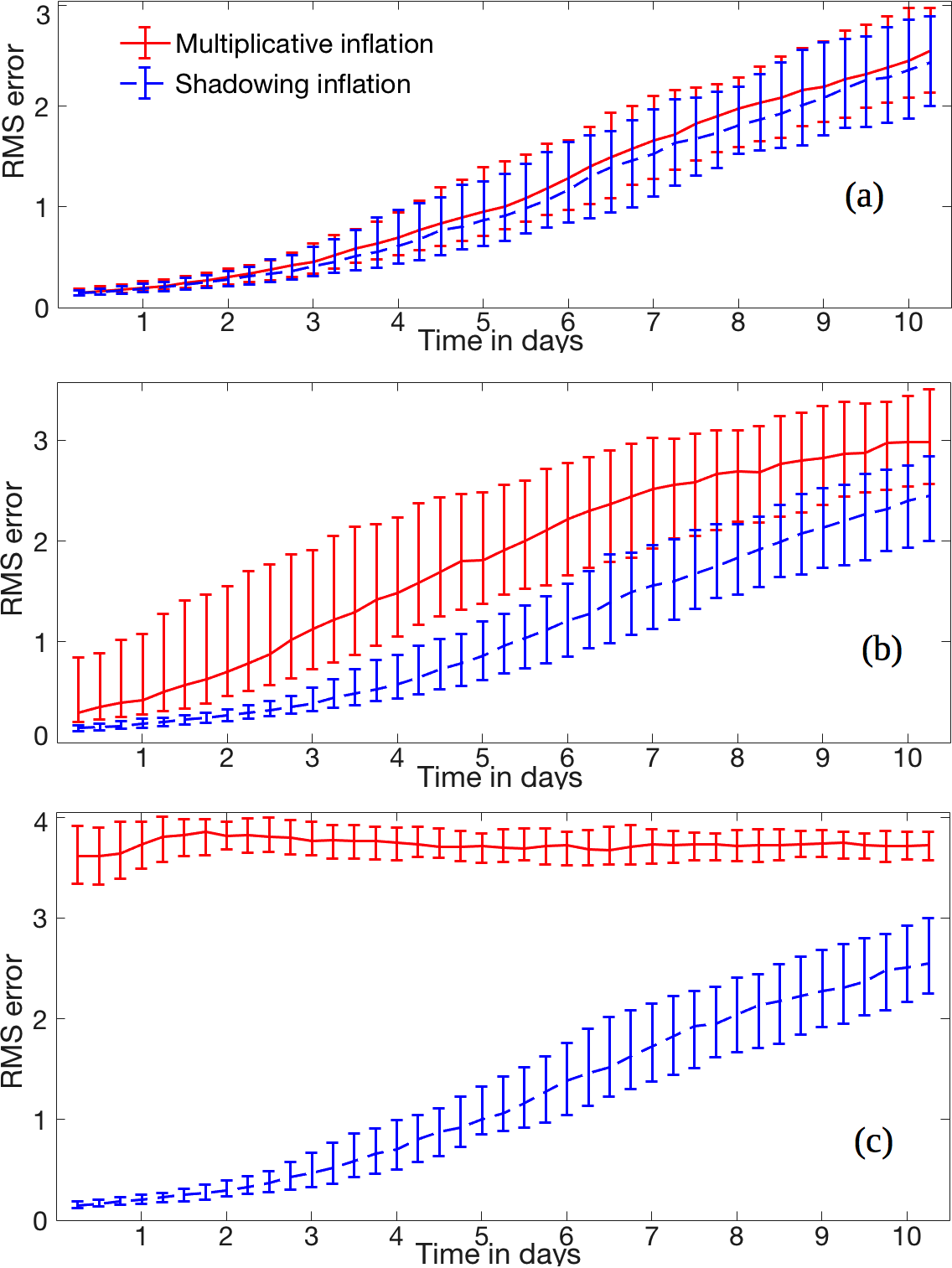}
\end{center}
\caption{This figure plots the median and interquartile RMS errors of all locations for the multiplicative inflation scheme (solid, red) and the shadowing inflation scheme (dashed, blue) for a ten day forecast at the end of the 110 day DA simulation. (a) is for a shadowing inflation factor of $\delta=0.02$, (b) is for a shadowing inflation factor of $\delta=0.05$, and (c) is for a shadowing inflation factor of $\delta=0.10$ (and a corresponding multiplicative inflation factor of $1+\delta$ in all cases). Here, the localization radius is $r=5$ and there are $8$ fixed observations.}
\label{forecastInf}
\end{figure}

Figure \ref{forecastInf} plots the median and interquartile RMS errors of 10 day forecasts performed after the 110 day DA simulation for both the multiplicative inflation scheme (solid, red) and the shadowing inflation scheme (dashed, blue). Figure \ref{forecastInf}a is for an inflation factor of $\delta=0.02$, Figure  \ref{forecastInf}b is for an inflation factor of $\delta=0.05$ (where this forecast begins at the last day of the simulation plotted in Figure \ref{fig:timeseries}a), and Figure \ref{forecastInf}c is for an inflation factor of $\delta=0.1$. In Figure \ref{forecastInf}a, both schemes perform similarly, with the shadowing scheme performing slightly better. For both Figures \ref{forecastInf}b and Figure \ref{forecastInf}c the forecasted solution, or shadowing time, of the shadowing inflation scheme outperforms the multiplicative inflation scheme; thus demonstrating the robustness of the shadowing scheme over a range of parameter values.

%As illustrated in Figure \ref{fig:sampleEnsembleTraj}, the shadowing inflation can lead to a better forecast, since it better spreads the ensemble of trajectories about the observed true trajectory. For Figures \ref{forecastInf}b and Figure \ref{forecastInf}c the forecasted solution, or shadowing time, of the shadowing inflation scheme substantially outperforms the multiplicative inflation scheme. 

\begin{figure}
\begin{center}
\includegraphics[width=\columnwidth]{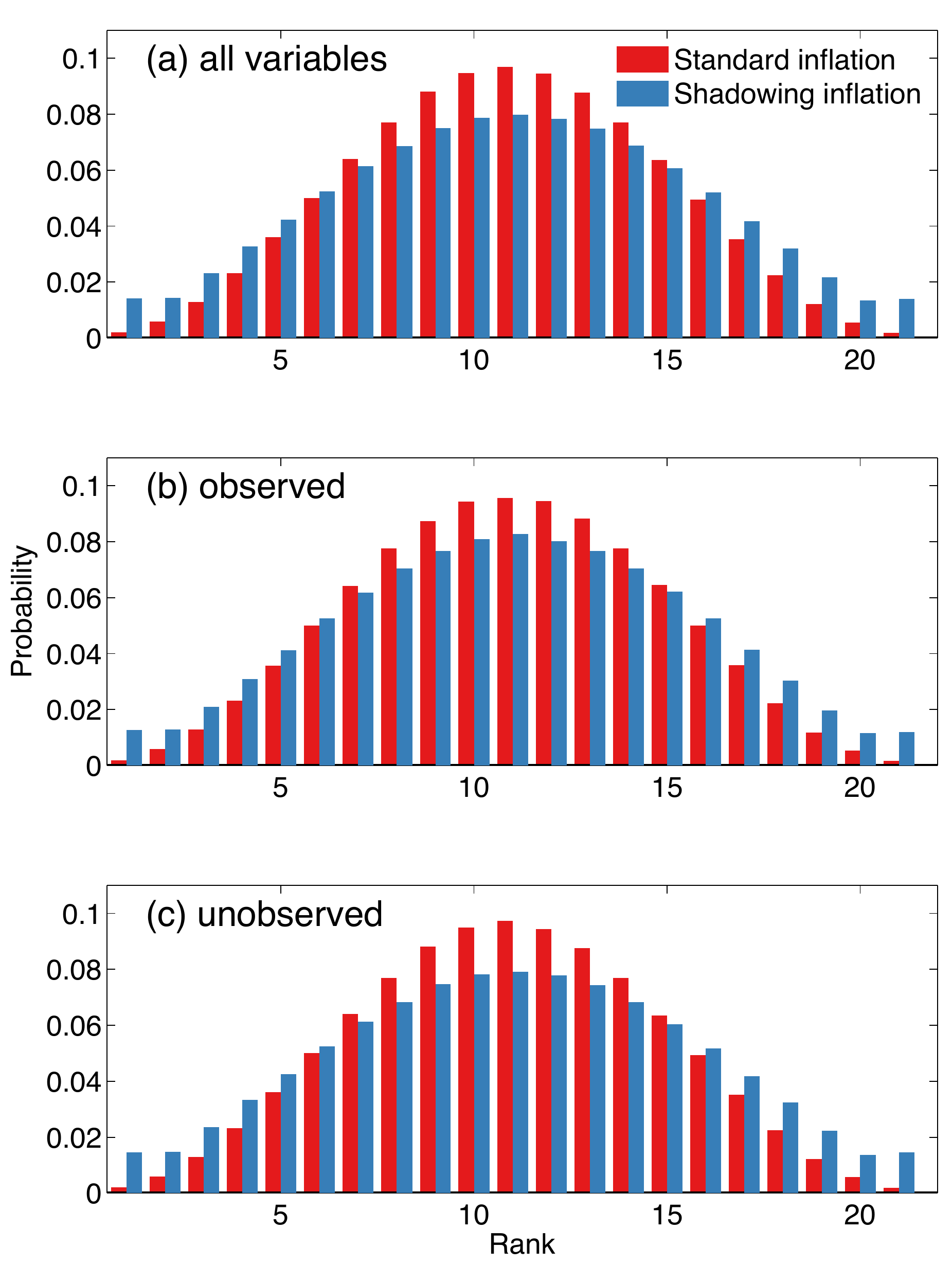}
\end{center}
\caption{Rank histograms, $N_{obs}=8$, $r=5$, $\delta = 0.02$.}
\label{fig:rankHist_N8_r2}
\end{figure}

Finally, we show that shadowing inflation improves ensemble reliability by comparing ranked probability histograms for the two schemes \cite{wilks2011statistical}.
Figure \ref{fig:rankHist_N8_r2} shows ranked probability histograms for the shadowing and standard methods for a shadowing inflation factor of  $\delta=0.02$.
The convex shape of the diagrams indicates that both methods are overdispersive,
however the shadowing inflation method is significantly less underconfident (across both observed and unobserved variables) than the standard multiplicative inflation method.

\section{Discussion}

This work has introduced a new shadowing-based inflation method for ensemble data assimilation. We have tested this shadowing inflation scheme numerically on a low-dimensional nonlinear system exhibiting chaotic dynamics reminiscent of those in the atmosphere.
Comparing shadowing inflation with standard global multiplicative covariance inflation, we have found that shadowing inflation outperforms the standard multiplicative inflation over a range of inflation values exhibiting a relative insensitivity to parameter tuning, often leading to longer forecast shadowing times, and maintaining ensemble reliability. 

All experiments for the present work were performed using a perfect model --
an obvious area for further exploration is the case of model error,
where ensemble inflation must compensate for structural deficiencies in the forecast model.
That shadowing inflation tends to perform best at the extremities of the attractor 
(as shown in Figure \ref{fig:sampleEnsembleTraj}) 
suggests that it might be a useful method in situations involving model error,
as it is near these extremes, or near a saddle point in the stable manifold, where model error should have a large detrimental effect.
Future work will involve coupling shadowing inflation methods with known methods for dealing with model error \cite{Mitchell2015}.

The present work also uses a constant inflation factor, regardless of where in state space the inflation is being performed.
However, as shown in \cite{Lieb-Lappen2012} for the forecasting problem,
it can be beneficial to perform adaptive inflation depending on the location of the forecast on the model attractor.
While we noticed no significant difference when making the inflation factor dependent on the level of expansion measured by the singular values $s_i$,
a more attractor-based scheme may show further improvements.
Note that such methods would necessarily be constrained to low-dimensional systems where the shape of the attractor can be reasonably estimated.
The shadowing inflation factor may also be related to the observational density.
The use of uniformly-spaced observations leads to large unobserved gaps for some values of $N_{\rm obs}$ (e.g.~$N_{\rm obs} = 7$); 
exploring the dependence of optimal inflation level on $N_{\rm obs}$ will require us to use a different observational setup to that employed here.

For simplicity we formulated the method here involving an extra SVD of the full forecast ensemble deviation matrix,
which would be computationally impractical in large systems.
Furthermore, analysis ensembles tend to be underdispersive in high-dimensional systems, requiring large inflation factors -- care must therefore be taken when performing inflation with such systems.
Significant computational savings could be made in high-dimensional systems by performing the shadowing inflation within the ensemble space.
%utilizing the SVD step which is already performed as part of the LETKF algorithm.
However, how to best implement this decomposition, particularly while ensuring fidelity and sufficient smoothness of the interpolated fields, remains a question for exploration.
Future work will formulate a shadowing inflation scheme within the ensemble subspace and compare the accuracy and computational cost when making this modification.

\section*{Acknowledgments}
The authors acknowledge and thank the Mathematics and Climate Research Network, in particular National Science Foundation support DMS-0940271 and DMS-0940314.

\section*{References}

\bibliography{references}

\end{document}